\documentclass[%
 reprint,
superscriptaddress,
 amsmath,amssymb,
 aps,pre
floatfix,
]{revtex4-1}
\usepackage{graphicx}
\usepackage{dcolumn}
\usepackage{bm}
\usepackage[colorlinks=true,linkcolor=blue,citecolor=blue]{hyperref}

\usepackage{amsmath,bm}
\DeclareMathAlphabet\mathbfcal{OMS}{cmsy}{b}{n}

\renewcommand{\u}{\bm{u}}
\renewcommand{\b}{\bm{b}}
\renewcommand{\r}{\bm{r}}
\newcommand{\x}{\bm{x}}
\renewcommand{\l}{\bm{\ell}}

\usepackage{float} 
\usepackage[normalem]{ulem}
\usepackage{xcolor}

\begin{document}

\title{Local cascade and dissipation in incompressible Hall magnetohydrodynamic turbulence: the Coarse-Graining approach}
\author{D. Manzini}
\affiliation{Laboratoire de Physique des Plasmas (LPP), CNRS, École Polytechnique, Sorbonne Université, Université Paris-Saclay, Observatoire de Paris, 91120 Palaiseau, France}%
\affiliation{Dipartimento di Fisica E.Fermi, University of Pisa, Italy}%

\author{F. Sahraoui}
\affiliation{Laboratoire de Physique des Plasmas (LPP), CNRS, École Polytechnique, Sorbonne Université, Université Paris-Saclay, Observatoire de Paris, 91120 Palaiseau, France}%

\author{F. Califano}
\affiliation{Dipartimento di Fisica E.Fermi, University of Pisa, Italy}%

\author{R. Ferrand}
\affiliation{Laboratoire de Physique des Plasmas (LPP), CNRS, École Polytechnique, Sorbonne Université, Université Paris-Saclay, Observatoire de Paris, 91120 Palaiseau, France}%

\begin{abstract}
We derive the coarse-graining (CG) equations of  incompressible Hall Magnetohydrodynamics (HMHD) turbulence to investigate the {\it local (in space)} energy cascade rate as a function of the filtering scale  $\ell$. 
First, the CG equations are space averaged to obtain the analytical expression of the mean cascade rate. Its application to 3 dimensional (3D) simulations of (weakly compressible) HMHD shows a cascade rate consistent with the value of the mean dissipation rate in the simulations and with the classical estimates based on the ``third-order" law. Furthermore, we developed an anisotropic version of CG that allows us to study the magnitude of the cascade rate along different directions with respect to the mean magnetic field. Its implementation on the numerical data with moderate background magnetic field shows a weaker cascade along the magnetic field than in the perpendicular plane, while an isotropic cascade is recovered in the absence of a background field. The strength of the CG approach is further revealed when considering the {\it local-in-space} energy transfer, which is shown theoretically and numerically to match at a given position $\x$, when locally averaged over a neighboring region, the (quasi-)local dissipation. Prospects of exploiting this new model to investigate local dissipation in spacecraft data are discussed. 
\end{abstract}
\maketitle
\section{Introduction} \label{sec:intro}


Turbulence plays a key role in space and astrophysical plasmas as, for instance, it mediates energy conversion stored on large-scale fields into particle heating and/or acceleration at smaller scales. The standard theory of turbulence for hydrodynamics predicts an energy cascade from large scales, where it is injected, to the small scales where it is dissipated by viscosity stemming from particle collisions at the microscopic level \citep{batchelor_theory_1953,frisch_turbulence_1995,monin_statistical_2013}.
On the other  hand, the low density and high temperature conditions of most heliospheric plasmas make them nearly collisionless. In those plasmas a turbulent cascade and particle heating are frequently observed, however, the precise mechanisms by which the turbulent fluctuations of the electromagnetic fields and plasma flow are damped still elude our full understanding \citep{bruno_solar_2013,matthaeus_who_2011,goldstein_kinetic_2015,sahraoui_magnetohydrodynamic_2020}. 

A key step in answering these fundamental questions is to identify and characterize the regions of plasma involved in intense cross-scale energy transfers. A popular tool that has been widely used in turbulence studies, in particular those based on spacecraft observations, is the so-called ``third-order" law: a statistical relation that links the mean energy cascade rate (equal in the formalism to the rate of energy injection and dissipation) to the turbulent fluctuations at a given scale. The theoretical models used range from Incompressible MHD \citep{politano_dynamical_1998} to more complex systems that involve density fluctuations and/or small (sub-ion) scale effects \citep{banerjee_exact_2013,andres_alternative_2017,andres_energy_2018,hellinger_von_2018, ferrand_exact_2019,ferrand_compact_2021}. All these studies have greatly helped to gain deeper insight into the turbulence dynamics in a variety of heliospheric plasmas, including the solar wind (SW) \citep{smith_dependence_2006,sorriso-valvo_observation_2007,marino_heating_2008,stawarz_turbulent_2010,osman_anisotropic_2011,macbride_turbulent_2008,coburn_third-moment_2015,banerjee_alternative_2016,hadid_energy_2017} and planetay magnetospheres \citep{hadid_compressible_2018,andres_solar_2020,sorriso-valvo_turbulence-driven_2019}. The ``third-order" law, albeit rigorous and derived under fairly non-restrictive hypotheses, does however require ensemble averages computed as time and/or space average when applied to simulations or spacecraft data under the assumption of ergodicity \citep{frisch_turbulence_1995}. As such, those laws fail to describe cross-scale energy transfer in localized regions of space. To overcome this shortcoming some heuristic tools have been proposed such as the Local Energy Transfer (LET) \citep{sorriso-valvo_turbulence-driven_2019}, which relaxes the statistical average used in the ``third-order" law, or the Partial Variance Index (PVI)  \citep{Greco2008, chasapis_thin_2015} used to localize regions of space with large magnetic shear, a proxy to identify regions of strong electric current. However, those tools lack a solid theoretical foundation, which is mandatory to justify their use as a means to measure energy rates. The present work fills this gap by providing a robust theoretical model based on filtering (or coarse-graining) the Hall-MHD (HMHD)  equations that retains spatial locality while allowing one to recover results consistent with the ``third-order" law once spatially averaged. We furthermore show analytically and numerically that the local (in space) energy cascade rate across a scale $\l$ is a good proxy to measure local dissipation within limited regions of space. 

Note that other local theories of turbulence based on the concept of inertial dissipation have been proposed in recent years to include the role of discontinuities in dissipating energy, which are not rigorously accounted for in the ``third-order" law formalism \citep{galtier_origin_2018,dubrulle_beyond_2019}. Those models have been recently used to compute dissipation within discontinuities observed in spacecraft data \citep{Kuzzay_PRE_19,david_proof_2021,david_energy_2022}

%
\section{Coarse-Grained Incompressible HMHD equations}\label{CG}
We introduce the notion of a coarse grained (CG) measurement of a field $f$ with a scale resolution $\ell$. We choose a kernel function $G$ normalized to one $\int d\r G(\r)=1$, centered $\int d\r \r G(\r)=0$ and with variance of order unity $\int d\r |\r|^2 G(\r)  \simeq 1$. For any given scale $\ell$ we define $G_\ell(\x)=\ell^{-3}G(\x/\ell)$ so that the normalization is pres\textbf{}erved but the variance is now of order $\ell^2$. The coarse-graining operation is defined as
\begin{equation}
\bar{f}_\ell(\x)=\int d\r G_\ell(\r)f(\x+\r).
\end{equation}
It represents a local average of $f$ on a spatial region of radius $\sim \ell$ centered around the point $\x$. This convolution smooths out the fluctuations with scale smaller than $\ell$, and gives a coarser representation of the field, hence the name. Coarse-Graining a field $f$ at a given scale $\ell$ individuates two different quantities: the large-scale field $\bar{f}_\ell$, which in virtue of the smoothing operation retains only the scales $>\ell$ (in $\mathbf{k}$-space the wave-vectors $|\bm{k}|\lesssim 1/\ell$) and the \emph{un-resolved} or \emph{sub-scale} field $f'=f-\bar{f}_\ell$ which accounts for scales $<\ell$ (wave-vectors $\bm{k}\gtrsim 1/\ell$). The term ``unresolved" comes from the fact that when coarse-graining the fields we choose to resolve fluctuations down to the scale $\ell$ only.
To keep the notation simple we will omit the subscript (filtering scale) $\ell$ when not strictly necessary and denote $\bar{f}_\ell$ simply as $\bar{f}$. The quality of the filtering in $\bm{k}$ space depends on the choice of the filtering function $G_\ell$. For instance, a sharp spectral filter such as $G_\ell^{sp}=\Pi_{i=1}^3sin(\pi x_i/\ell)/(\pi x_i)$ allows one to clearly separate between scales but it looses the spatial locality, while a box filter  $G^{box}_\ell=\ell^{-3}, \; |x_i|<\ell/2 $,  $G^{box}_\ell=0, \; |x_i|\geq\ell/2$ is local in space, but does not allow for unambiguous separation between scales. Other filters with intermediate properties can be defined such as the Gaussian filter $G^{Gss}_\ell=(2/\pi\ell^2)^{3/2}exp\{-2|\x|^2/\ell^2\}$ used in this work \citep{meneveau_scale-invariance_2000, eyink_localness_2009}.\\ 
We start from the incompressible Hall-MHD (IHMHD) equations normalized to Alfv\'en units:
\begin{equation}
\begin{split}
    &\partial_t \u=-(\u\cdot\nabla)\u+(\b\cdot\nabla)\b-\nabla P +\bm{d}_\nu+\bm{f} \\
    &\partial_t\b= \nabla\times (\u\times\b)-d_i\nabla\times(\bm{j}\times\b)+\bm{d}_\eta \\
    &\nabla\cdot\u=0
    \end{split}
\label{eq:HMHD}
\end{equation}
where $\b={\bf B}/\sqrt{\mu_0\rho_0}$ is the (scale dependent) Alfv\'en speed, $\rho_0$ the mean plasma density, ${\bm{j}=\nabla\times\b}$, $d_i$ is the ion inertial length, $P=p/\rho_0+|\b|^2/2$ is the total pressure, $\bm{d}_\nu, \bm{d}_\eta$ are the velocity and magnetic field dissipation terms, respectively, and $\bm{f}$ is an external force injecting energy at large scales. The CG operation is a convolution and therefore commutes with space and time derivatives. The equations filtered at scale $\ell$ are readily obtained by convolving  equations \eqref{eq:HMHD} with the filtering kernel $G_\ell$
\small
 \begin{subequations}
     \begin{gather}
     \label{eq:MHD_filU}
    \partial_t\bar \u=-(\bar{\u}\cdot\nabla)\bar{\u}+(\bar{\b}\cdot\nabla)\bar{\b}-\nabla\bar{P}+\bar{\bm{d}_{\nu}}-\nabla\cdot \bm{\tau}+\bar{\bm{f}}\\
    \label{eq:MHD_filB}
    \partial_t\bar{\b}=\nabla\times\left[(\bar{\u}-d_i\bar{\bm{j}})\times\bar{\b}\right]+\bar{\bm{d}}_\eta+\nabla\times(\mathbfcal{E}_\text{MHD}+d_i\mathbfcal{E}_{\text{Hall}})
\end{gather}
 \end{subequations}
 \normalsize
where we introduced the notations $\tau(f,g)=\overline{fg}-\bar{f}\bar{g}$ and, for the sake of readability, we define the second order tensor $\tau_{ij}=\tau(u_i,u_j)-\tau(b_i,b_j)=\overline{u_iu_j}-\bar{u}_i\bar{u}_j-\left(\overline{b_ib_j}-\bar{b}_i\bar{b}_j \right)$.\\

These equations describe the dynamics of the CG (large scale) fields and closely resemble the HMHD equations \eqref{eq:HMHD}. The difference lies in the presence of additional contributions stemming from the filtering of the nonlinear terms. These quantities represent the action of the ``unresolved" scales ($ < \ell$) on the filtered fields. In particular, in equation \eqref{eq:MHD_filU} we find the divergence of the subscale Reynolds and Maxwell stress tensors $\tau_{ij}=\tau(u_i,u_j)-\tau(b_i,b_j)$,  while in equation \eqref{eq:MHD_filB} we find the curl of the subscale electric field in the MHD limit $\mathbfcal{E}_\text{MHD}=\overline{\u\times\b}-\bar{\u}\times\bar{\b}$ and the correction due to the Hall term $\mathbfcal{E}_{\text{Hall}}=-\left (\overline{\bm{j}\times\b}-\bar{\bm{j}}\times\bar{\b}\right )$. 

Multiplying equation \eqref{eq:MHD_filU} by $\bar{\u}$ and equation \eqref{eq:MHD_filB} by $\bar{\b}$ we obtain the time evolution of the large scale kinetic and magnetic energy densities:
 \small
 \begin{subequations}
    \begin{align}
    \label{eq:MHD_en_filU}
        \begin{split}
            \partial_t \frac{|\bar\u|^2}{2}=&-\nabla\cdot\left[ \frac{|\bar{\u}|^2}{2}\bar{\u} + \bar{P}\bar{\u}-(\bar{\u}\cdot\bar{\b})\bar{\b} +\bm{\tau}\cdot\bar{\u}\right]\\-&\bar{\b}\cdot(\bar{\b}\cdot\nabla)\bar{\u}+\bar{\u}\cdot\bar{\bm{d}_\nu}-\pi^u +\bar{\u}\cdot\bar{\bm{f}}
        \end{split}\\
    \label{eq:MHD_en_filB}
        \begin{split}
            \partial_t \frac{|\bar{\b}|^2}{2}=&-\nabla\left[ \frac{|\bar{\b}|^2}{2}\bar{\u}+d_i(\bar{\bm{j}}\times\bar{\b})\times\bar{\b}-(\mathbfcal{E}_\text{MHD}+d_i\mathbfcal{E}_{\text{Hall}})\times\bar{\b}\right]\\&+\bar{\b}\cdot(\bar{\b}\cdot\nabla)\bar{\u}+\bar{\b}\cdot\bar{\bm{d}_\eta}-\pi^{b,\text{MHD}}-d_i\pi^{b,\text{Hall}}
        \end{split}
    \end{align}
\label{eq:MHD_en_fil}
\end{subequations}
 \normalsize where we introduced the quantities:
 \begin{equation}
 \pi^u=-\partial_j\bar{u}_i\tau_{ij} \quad \pi^{b,\text{MHD}}=-\bar{\bm{j}}\cdot\mathbfcal{E}_{\text{MHD}}\quad \pi^{b,\text{Hall}}=-\bar{\bm{j}}\cdot\mathbfcal{E}_\text{Hall}.
 \label{eq:pi_def}
\end{equation} 
These terms, the study of which is the main focus of this work, are the local (in space) energy transfers across the scale $\ell$. They appear as a sink in the large scale energy equations \eqref{eq:MHD_en_fil} and a source in the small scales ones (see Appendix \ref{ap:sm_sc}).
 
 While these equations allow us to analyze separately the magnetic and kinetic energy cascades, here we rather focus on the study of the cascade of the total energy (note that in the energy balance equation the internal energy is not included as it is a conserved quantity in incompressible pressure-isotropic flows \citep{simon_general_2021}). Summing equations \eqref{eq:MHD_en_filU}-\eqref{eq:MHD_en_filB} and separating the cascade rate into its MHD component, $\pi^{\text{MHD}}=\pi^u+\pi^{b,\text{MHD}}$, and Hall component $\pi^{b,Hall}$, hereafter denoted simply as $\pi^{\text{Hall}}$, we find:
\begin{equation}
 \begin{split}
     &\partial_t\left(\frac{|\bar{\u}|^2+|\bar{\b}|^2}{2}\right)\\+&\nabla\cdot\left[ \frac{|\bar{\u}|^2+|\bar\b|^2}{2}\bar{\u}+\bar{P}\bar{\u}-(\bar{\u}\cdot\bar{\b})\bar{\b}+d_i(\bar{\bm{j}}\times\bar{\b})\times\bar{\b}\right]\\
     +&\nabla\cdot\left[\bm\tau\cdot\bar{\u}- (\mathbfcal{E}_I+d_i\mathbfcal{E}_H)\times\bar{\b}\right]\\-&\bar{\u}\cdot\bar{\bm{d}_\nu}-\bar{\b}\cdot\bar{\bm{d}_\eta}-\bar{\u}\cdot\bar{\bm{f}}=\\-&\pi^\text{MHD}-d_i\pi^\text{Hall}
\end{split}
\label{eq:MHD_en_fil_tot}
 \end{equation}
This equation is the starting point of our study and the first rigorous result of this paper, which extends previous results \cite{aluie_coarse-grained_2017} to IHMHD. Equation (\ref{eq:MHD_en_fil_tot}) shows that, at a given position $\x$, the time variation of the large scale energy density is a balance between the spatial advection due to large and small scale fields  (the second and third lines, respectively), the effects of dissipation and forcing at large scales (fourth line) and, lastly, the energy transfer across the scale $\ell$, namely $\pi^{\text{MHD}}$ and $d_i\pi^{\text{Hall}}$. We recall that in equation (\ref{eq:MHD_en_fil_tot}) the filtering scale $\ell$ can be varied to gauge the magnitude of each term as a function of scale and its choice individuates two separate range of the spectrum: the resolved large scales (corresponding to wavevectors $|\bm{k}|\lesssim 1/\ell$) and the ``unresolved" small scales ($|\bm{k}|\gtrsim 1/\ell$).

 \section{Space Integration and the Cascade Rate}\label{sec:global}
 Equation \eqref{eq:MHD_en_fil_tot} describes the temporal evolution of the large scale energy density. By performing a spatial average over the whole domain, denoted $\langle \rangle$, and assuming no net flux at the boundaries, we recover the following expression for the temporal evolution of the mean large scale energy $\bar{E}=\left\langle|\bar{\u}|^2+|\bar{\b}|^2\right\rangle$/2,   
\begin{equation}
     \partial_t \bar{E}=-\Pi+\langle \bar{\u}\cdot\bar{\bm{d}}_\nu+\bar{\b}\cdot\bar{\bm{d}}_\eta\rangle + \langle\bar{\u}\cdot\bar{\bm{f}}\rangle
     \label{eq:En_LS_int}
 \end{equation}
 where we introduced $\Pi = \Pi^\text{IMHD} + d_i\Pi^\text{Hall} = \langle \pi^\text{IMHD}\rangle + d_i\langle\pi^\text{Hall} \rangle$.
 
 All the quantities involved in Eq.\eqref{eq:En_LS_int} are functions of the filtering scale $\ell$ only, as the spatial dependence is lost when averaging over the spatial domain.
 We see that the large scale energy is affected by three processes: the forcing mechanism that injects energy via the term  $\langle \bar{\u}\cdot\bar{\bm{f}}\rangle$, the effect of dissipation at scales larger than the filtering scale $\ell$ given by the term $\langle \bar{\u}\cdot\bar{\bm{d}}_\nu+\bar{\b}\cdot\bar{\bm{d}}_\eta\rangle$ and, lastly, the large scale energy transfer due to nonlinearities given by  $\Pi(\ell)=-\partial_t\bar{E}|_\text{NL}$, where we denote with $\partial_t(\cdot)|_\text{NL}$ the rate of change due to non linear processes. $\Pi(\ell)$ stands for the cross-scale energy transfer rate, that is: the amount of energy flowing from the large-scale, resolved quantities ( $>\ell$) to the small-scales ``unresolved" ones ($<\ell$). In Fourier space it writes
 \begin{equation}
     \Pi(\ell)=-\frac{\partial}{\partial t} \left[(2\pi)^3\int d\bm{k} \left(\frac{|\hat{\u}(\bm{k})|^2+|\hat{\b}(\bm{k})|^2}{2}\right)|\hat{G}_\ell(\bm{k})|^2 \right]_\text{NL}
    \label{eq:PI}
 \end{equation}
 where the squared modulus of the filtering function $|\hat{G}_\ell(\bm{k})|^2$ plays the role of a low-pass filter. Eq.(\ref{eq:PI}) shows that $\Pi$ closely resembles the formal definition of the cascade rate across the wave-vector $K=1/\ell$ \citep{frisch_turbulence_1995}:
\begin{equation}
    -\frac{\partial}{\partial t} \left[(2\pi)^3\int_{|\bm{k}|\leq K} d\bm{k} \left(\frac{|\hat{\u}(\bm{k})|^2+|\hat{\b}(\bm{k})|^2}{2}\right) \right]_\text{NL}
\end{equation}
if one replaces the sharp cut-off in ${\bf k}$-space with a gentler slope given by the shape of $|\hat{G}_\ell(\bm{k})|^2$. In line with this view, for a given (short) interval $\Delta t$, the variation of the large scale energy due to nonlinear processes is $-\Pi(\ell)\Delta t$. This energy is directly transferred to  small scales $<\ell$ so that a positive $\Pi(\ell)$ is the signature of a direct cascade with energy flowing from large to small scales, while a negative value yields an inverse cascade (i.e., from small to large scales).

It is our understanding that the CG quantity $\Pi$ represents an extension over the current state of the art ``third-order" laws that, under a set of assumptions, provide an estimate of the cascade rate, often denoted $\varepsilon$. Indeed ``third order" laws rely heavily on the assumption of the existence of an inertial range and on the ergodicity hypothesis to compute ensemble averages as time and/or space averages, while equation \eqref{eq:MHD_en_fil_tot} holds without the need of those assumptions. Thus, it allows us to estimate the transfer rate at any scale, i.e. not necessarily in the inertial range, and {\it regardless} of how important is dissipation \citep{eyink_cascades_2018-1}. This remark might prove to be useful in particular in collisionless plasmas where dissipation (via, e.g., Landau damping) may occur at all scales, which would question the very existence of the inertial range \citep{ferrand_fluid_2021}.

Nevertheless, if Kolmogorov hypotheses are satisfied we expect the two quantities $\Pi$ and $\varepsilon$ to converge to the same value. This can be readily explained by recalling the formal definition of $\varepsilon$ as the time evolution of the auto-correlation function due to nonlinear terms \citep{frisch_turbulence_1995}:
 \begin{equation}
     \varepsilon(\bm{l})=-\frac{1}{2}\partial_t\langle\u(\x)\cdot\u(\x+\bm{l})+\b(\x)\cdot\b(\x+\bm{l})\rangle|_\text{NL}
     \label{eq:def_eps}
 \end{equation}
 where $\langle \rangle$ denotes an ensemble average, which is computed as a space average under the assumption of ergodicity.
 Using Parseval theorem we can recover from the definition of $\varepsilon$, the expression:
 \begin{equation}
     \varepsilon(\bm{l})=-\frac{\partial}{\partial t} \left[ (2\pi)^3\int d\bm{q} \left( \frac{|\hat{\u}(\bm{k})^2|+|\hat{\b}(\bm{k})^2|}{2}\right)e^{-i\bm{q}\cdot\bm{l}}  \right]_\text{NL} 
 \end{equation}
 and performing a Fourier transform we obtain:
 \begin{equation}
      \int d\bm{k}\varepsilon(\bm{l})e^{i\bm{k}\cdot\bm{l}}=-(2\pi)^3\frac{\partial}{\partial t} \left[  \frac{|\hat{\u}(\bm{k})^2|+|\hat{\b}(\bm{k})^2|}{2}\right]_\text{NL}.
      \label{eq:epsilonFT}
 \end{equation}
Lastly, substitution of relation \eqref{eq:epsilonFT} in the definition of $\Pi$ given by Eq.\eqref{eq:PI} yields:
 \begin{equation}
    \Pi(\ell)=\int d\bm{l}' \varepsilon(\bm{l}')\phi_\ell(\bm{l}')
    \label{eq:PiFt}
\end{equation}
where the function $\phi_\ell(\bm{l}')=(2\pi)^{-3}\int d\bm{k} |\hat{G}_\ell(\bm{k})|^2 e^{i\bm{k}\cdot\bm{l}'}$ is directly related to the shape of the filtering function $G_\ell$. For the Gaussian filter used in this work, we have $\phi_\ell(\bm{l}')=(\pi\ell^2)^{-3/2}\exp\{-|\bm{l}'|^2/\ell^2\}$ so that the two quantities $\Pi$ and $\varepsilon$ differ by a Gaussian smoothing operation. It appears therefore that when Kolmogorov hypotheses are verified (and as such $\varepsilon$ is  constant in the inertial range) the two quantities have the same value and $\varepsilon$ is a good estimate of the cascade rate $\Pi$. 
\subsection{Numerical Validation}
\begin{figure}
\includegraphics[width=0.45\textwidth]{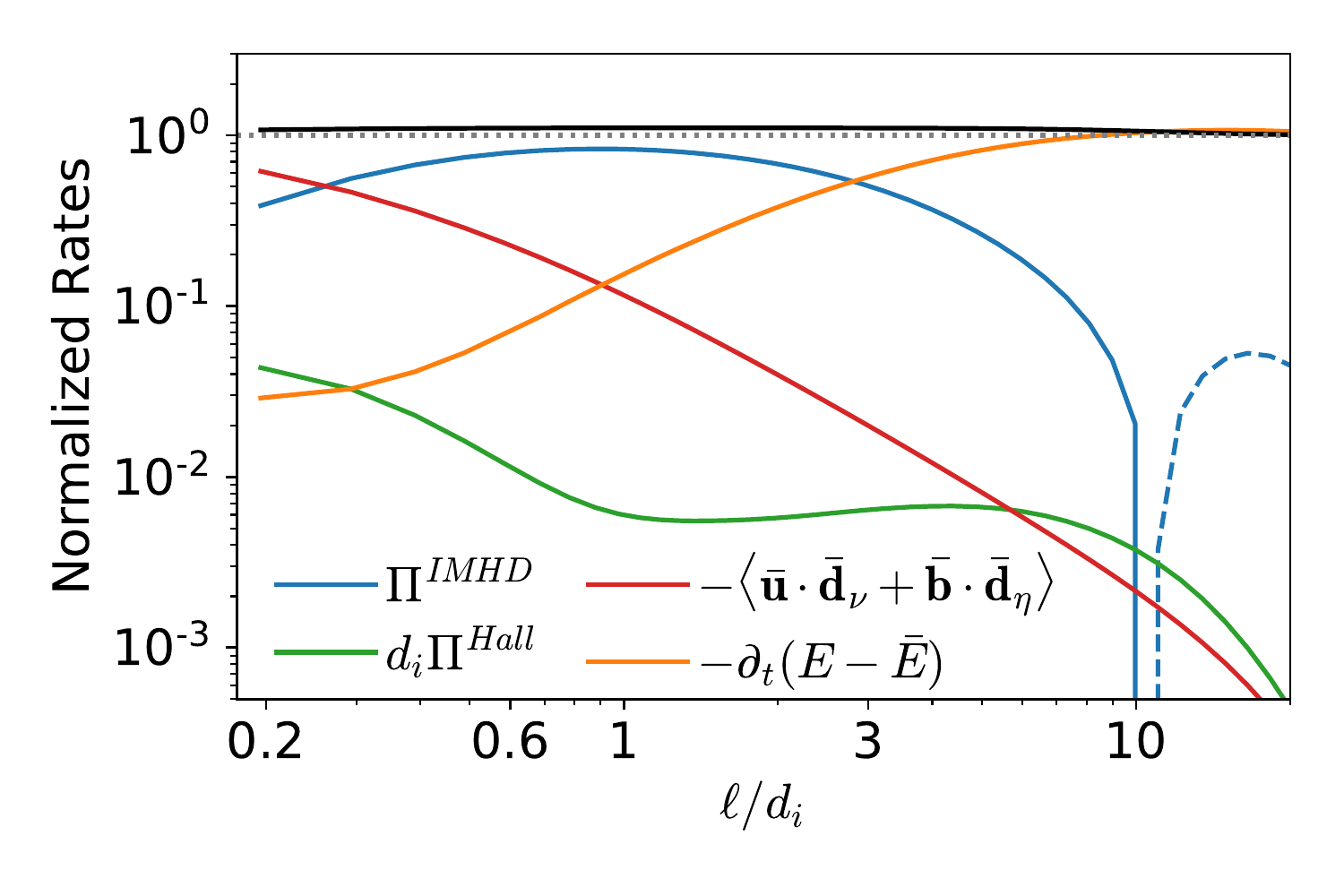}
\caption{Evaluation of the terms in Eq. \eqref{eq:sp_int_v2} for run I ($B=0$) normalized to the mean dissipation rate.
Dashed lines indicate negative values while the dotted horizontal line shows the reference value $\varepsilon/\varepsilon_{diss}=1$. The black line represents the sum of all the terms.}
\label{fig:validation_LS_int}
\end{figure}

In this section we will apply the theoretical results obtained above to simulation data that feature 3D freely-decaying weakly compressible HMHD  turbulence  (see \cite{ferrand_-depth_2022}). The two simulations were performed using the Fourier pseudo-spectral code GHOST \citep{gomez_mhd_2005,mininni_hybrid_2011}  on a $N=1024^3$ grid, spanning a real space cubic domain of side $L_0=50\times 2\pi d_i$ and a grid spacing of $\Delta\sim 0.05\times 2\pi d_i $. The two simulations were performed with two different values of the background field, namely $B_0=0$ (hereafter, run I) and $B_0=2$ (run II), which is very convenient to test the anisotropic CG approach that we introduce further below. In both runs dissipation is implemented via viscous and resistive terms and the dimensionless viscosity and magnetic diffusivity are taken to be equal $\nu=\eta=3\times10^{-4}$.  
The data cubes were taken when turbulence was deemed to have reached a fully developed state.\\
First, we estimate all terms of equation \eqref{eq:En_LS_int} using the simulation data of run I ($B_0=0$). To improve the graphical representation we add and subtract to the left side of equation \eqref{eq:En_LS_int} $-\partial_t E$, which by definition in freely decaying turbulence is $\partial_t E=-\varepsilon_{diss}$, so that we can write:
\begin{equation}
  \varepsilon_{diss} = - \partial_t(E-\bar{E}) + \Pi - \langle\bar{u} \cdot \bar{\bm{d}_\nu} + \bar{b} \cdot \bar{\bm{d}_\eta} \rangle
\label{eq:sp_int_v2}
\end{equation}
where we did not include the term $\langle\bar{\u}\cdot\bar{\bm{f}}\rangle$ as there is no forcing in our simulations.  The different terms of Eq. \eqref{eq:sp_int_v2} are shown in Fig.\ref{fig:validation_LS_int}.
We observe that the cascade rate (blue curve) in the central region $(0.3\lesssim\ell\lesssim 3)$ is representative of the dissipation rate. As $\ell$ tends to zero the effects of dissipation (red curve) on the large scales, \emph{i.e.} larger than the filtering scale $\ell$, become comparable to the total dissipation rate. Furthermore, the contribution of the Hall term to the energy cascade starts to increase around $\ell\sim d_i$ however, it remains negligible compared to the Ideal MHD contribution. This could be due to the small scale separation between the dissipative scales and $d_i$ \citep{ferrand_-depth_2022}.  The black line, which is the sum of the four terms displayed, remains 
constant within $10\%$ of the total dissipation rate for all values of $\ell$.

\subsection{Link with the ``third-order" law of IHMHD}\label{vkh}
We now turn to the assertion that $\Pi$ matches the classical "third-order" law theory of $\varepsilon$ when the latter is applicable. The standard approach to obtain an expression for $\varepsilon$ is to derive a generalized von K\'arm\'an–Howarth (vKH) equation for IHMHD turbulence  \citep{banerjee_alternative_2017, hellinger_von_2018,ferrand_exact_2019}, which describes the time evolution of the spatial auto-correlation function $R_E=\langle \u(\x)\cdot\u(\x+\bm{l})+\b(\x)\cdot\b(\x+\bm{l})\rangle/2$ where now the brackets $\langle \rangle$ denote formally an \emph{ensemble average}. This dynamical equation has contribution from both linear processes, such as the forcing or dissipation mechanisms and nonlinear ones. The latter are of particular interest as they provide the cross-scale interactions needed to sustain the energy cascade. The key quantity under study is the contribution to the rate of change $\partial_t R_E$ stemming from nonlinear processes. This quantity is denoted as $\varepsilon(\bm{l})=-\partial_t R_E(\bm{l})|_\text{NL}$. In the present work we use the vKH equation of IHMHD turbulence derived in \cite{banerjee_alternative_2017} (hereafter BG17). Denoting quantities evaluated at $\x+\bm{l}$ with a prime (\emph{e.g.} $\u(\x+\bm{l})=\u'$) and defining the field increments as $\delta \u=\u(\x+\bm{l})-\u(\x)=\u'-\u$, the BG17 law reads:  

 \begin{equation}\label{bg13}
 \begin{split}
     & \partial_t R_E(\bm{l})=\\
     &-\frac{1}{2}\langle\delta[\u\times\bm{w}+\bm{j}\times\b]\cdot\delta\u \rangle-\frac{1}{2}\langle\delta[(\u-d_i\bm{j})\times\b]\cdot\delta\bm{j}\rangle\\
     &+\frac{1}{2}\langle(\u'\cdot\bm{d}_\nu+\u\cdot\bm{d}'_\nu)+ (\u'\cdot\bm{d}_\eta+\u\cdot\bm{d}'_\eta)\rangle\\
     &+\frac{1}{2}\langle\u\cdot\bm{f}'+\u'\cdot\bm{f}\rangle
     \end{split}
 \end{equation}
  where we can identify the effects of non linearities (second line) which can be split into the MHD component from the Hall one (proportional to $d_i$), $\partial_t R_E|_{NL}=-[\varepsilon_\text{IMHD}+d_i\varepsilon_{\text{Hall}}]$, in addition to the dissipation and forcing terms (third and fourth line, respectively), which are simply denoted $\mathcal{D}, \mathcal{F}$ in the following. 
 
\begin{figure}
    \centering
    \includegraphics[width=0.45\textwidth]{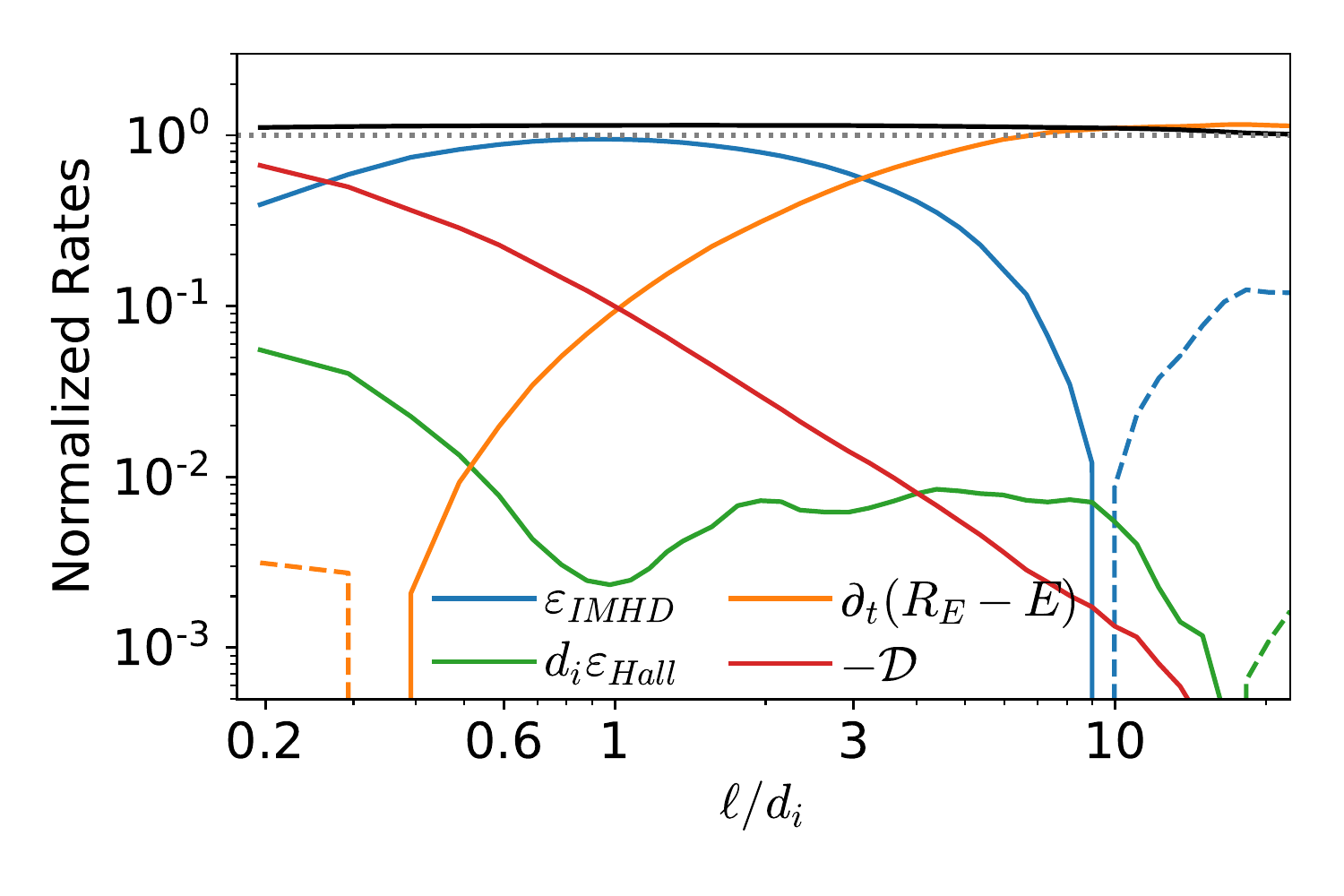}
    \caption{Numerical verification of the BG17 Law, the energy rates normalized to the mean dissipation rate. Negative values are represented with dashed lines, while the dotted horizontal line signals $\varepsilon/\varepsilon_{diss}=1$. The black line represents the sum of all the terms.  }
    \label{fig:EL_BG17}
\end{figure}
\begin{figure}[H]
    \centering
    \includegraphics[width=0.45\textwidth]{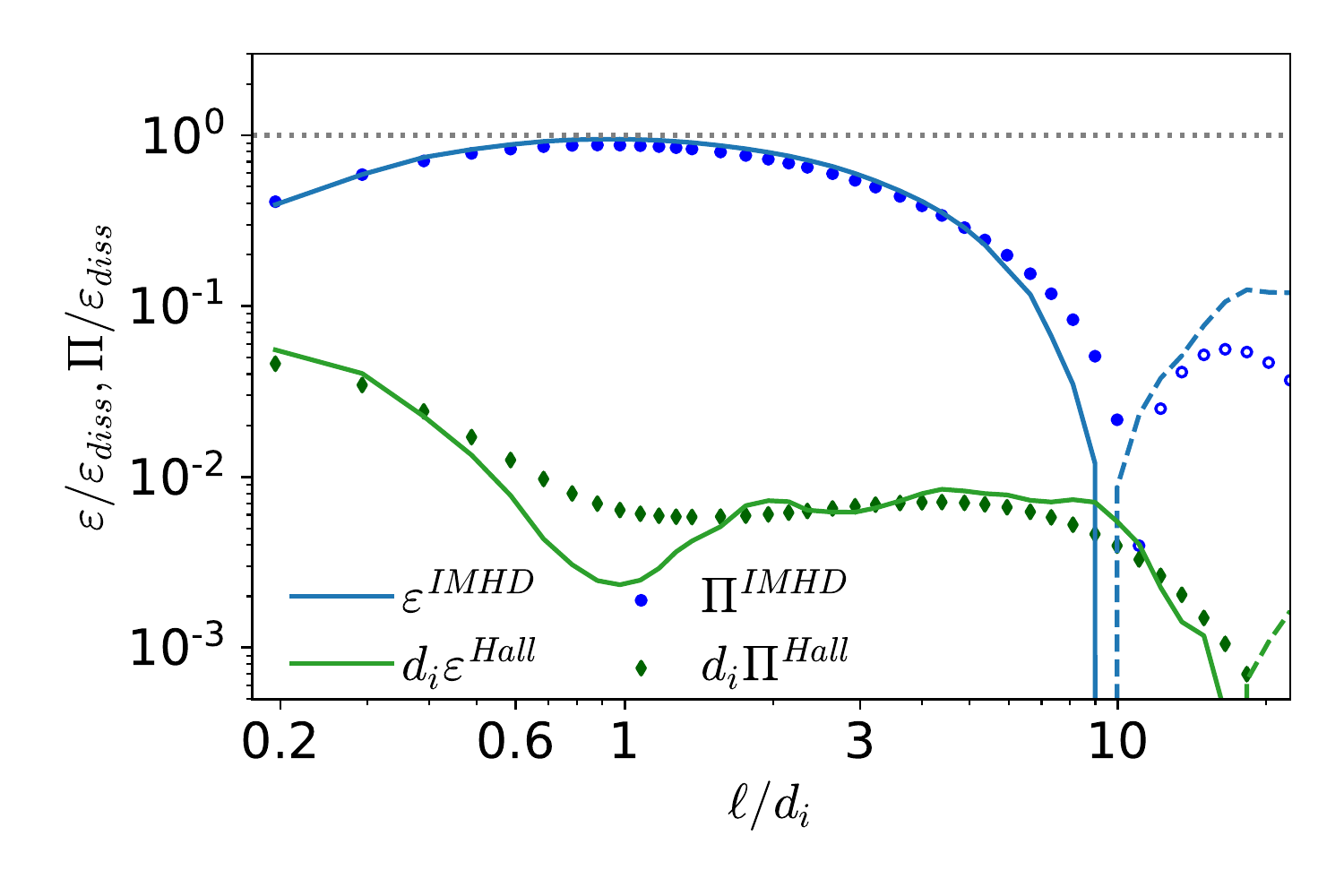}
    \caption{Comparison between the cascade rates $\Pi$ and $\varepsilon$ (BG17 law) for run I. The negative values are shown as dashed lines for $\varepsilon$ and as empty markers for $\Pi$.}
    \label{fig:EL_cmp}
\end{figure}

We evaluated all the terms in equation \eqref{bg13} using the data from run I, but the term $\mathcal{F}$ that does not apply here because of the free-decay nature of our simulations. Furthermore to improve the graphical representation we add and subtract to the left side of \eqref{bg13} $-\partial_t E$ which by definition, in freely decaying turbulence simulations, corresponds to the total dissipation rate $\partial_t E=-\varepsilon_{diss}$. We can re-write equation \eqref{bg13} in a more compact form:
\begin{equation}
    \varepsilon_{diss}=\partial_t(R_E-E)+\varepsilon_\text{IMHD}+d_i\varepsilon_\text{Hall}-\mathcal{D}
\end{equation}
The right hand side terms of this equation are plotted in Fig.\ref{fig:EL_BG17} where the spatial lag $\bm{l}$ is directly related to the scale $\l=\bm{l}/(2\pi)$.\\
The results in Fig.\ref{fig:EL_BG17} show that the sum of the different terms remains constant at all scales.  We observe a good resemblance between the cascade rates given by our CG model (see Fig.\ref{fig:validation_LS_int}) and those given by the ``third-order" law of IHMHD (Fig.\ref{fig:EL_BG17}). This is better emphasized when looking at Fig.\ref{fig:EL_cmp} where the various quantities are compared. 
\section {Anisotropic Cascade Rate}
In the presence of a background magnetic field $B_0 \neq0$ plasma turbulence becomes anisotropic \citep{shebalin_anisotropy_1983,matthaeus_particle_1984,goldreich_toward_1995,schekochihin_astrophysical_2009}, that is energy cascades preferentially in the direction perpendicular to the mean magnetic field. Turbulence anisotropy has been essentially investigated by looking at the energy spectra in $(k_\parallel,k_\perp)$ or, equivalently, at the second-order structure functions \citep{cho_2000,sahraoui_anisotropic_2006,meyrand_anomalous_2013,sahraoui_three_2010,chen2010}. Here we are interested in analyzing directly the anisotropy of the cascade rate. It is therefore mandatory to extend the CG approach beyond an isotropic treatment. Instead of using a spherically symmetric filtering kernel $G_\ell(\x)$ we can define a more general filter $G_{\l} (\x)$ that has different characteristic widths $\ell_x,\ell_y,\ell_z$ in the three real space directions. The CG quantity $\bar{f}=f\ast G_{\l} (\x)$ is low-pass filtered in an anisotropic way aimed at retaining mainly wave-vectors $k_x\lesssim 1/\ell_x,k_y\lesssim 1/\ell_y,k_z \lesssim1/\ell_z $. In this way we can highlight possible presence of plasma anisotropy. In the limit case when the filtering scales along two directions go to zero, \emph{e.g.} $\ell_x,\ell_z\to 0$, computing the quantity $\Pi$ allows one to recover the rate of energy flowing from $k_y\lesssim 1/\ell_y$ to $k_y\gtrsim 1/\ell_y$, thus recovering the 1-D cascade rate in the $y$ direction. Continuing our example, we choose a filtering function $G_{\l}(\x)=\delta(x)\delta(z)\psi_{\ell_y}(y)$, where $\delta$ and $\psi_{\ell_y}$ are respectively the Dirac and a 1-D filtering kernel with a characteristic width $\ell_y$. Then, the quantity $\Pi=-\partial_t\bar{E}|_\text{NL}$ can be written as:
\footnotesize
 \begin{equation*}
     \Pi_{\l} =-\frac{\partial}{\partial t}\left[(2\pi)^3\int dk_xdk_z\int dk_y \left(\frac{|\hat{\u}(\bm{k})|^2+|\hat{\b}(\bm{k})|^2}{2}\right) |\hat{\psi}_{\ell_y}(k_y)|^2\right]_{\text{NL}}
 \end{equation*}
 \normalsize
corresponding to the one-dimensional cascade rate along direction $y$. 
 
\begin{figure}[]
     \centering
     \includegraphics[width=0.45\textwidth]{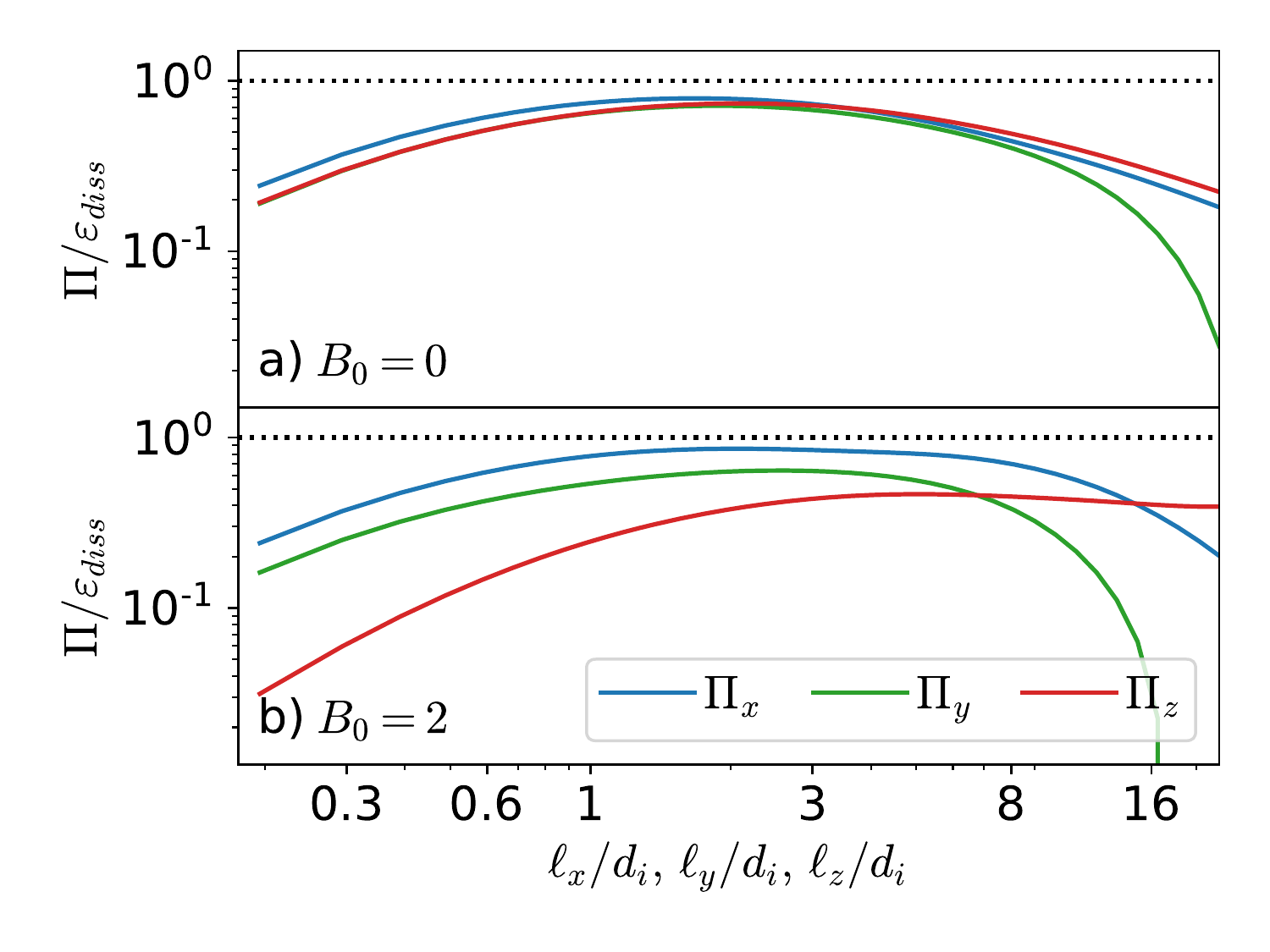}
     
     \caption{Anisotropic cascade rate for run I ($B_0=0$, top) and run I ($B_0=2$, bottom) along the three space directions $x,y,z$. A weaker cascade is evidenced in the $z$ direction when $B_0=2$.}
     \label{fig:Anisotropic}
 \end{figure}

The designed scheme is implemented on the two sets of simulations data obtained from run I and run II and the results are shown in Fig.\ref{fig:Anisotropic}.
We found that in the absence of a strong mean field the cascade rate is almost isotropic at most scales, with only  weak predominance of the cascade in the direction $x$ at the smallest scales. In the presence of a background field  (run II, $B_0=2$)  the cascade rate in the parallel direction is weaker in comparison with the two perpendicular directions, which look overall similar (but at the largest scales). In both runs, the violation of gyrotropy at the largest scales is likely to be a residual effect of the initial modes used to inject energy at largest scales of the simulation box.\\

\begin{figure}[]
     \centering
     \includegraphics[width=0.5\textwidth]{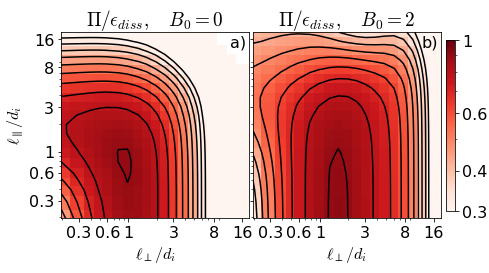}
     \caption{2D Cascade rate $\Pi(\ell_\perp,\ell_\parallel)$ normalized to the dissipation rate for run I ($B_0=0$, left) and run II ($B_0=2$, right).}
     \label{fig:my_anisotropic2D}
     \
 \end{figure}

A global picture of how energy flows in the  $(\ell_\parallel,\ell_\perp)$ plane can be obtained by computing the cascade rate $\Pi(\ell_\perp,\ell_\parallel)$, which measures the amount of energy that goes across wavevectors $|\bm{k}_\perp|\sim 1/\ell_\perp, k_\parallel=1/\ell_\parallel$. This is achieved by using the filtering function $G_{\ell_\perp,\ell_\parallel}=\varphi_{\ell_\perp}(x,y)\psi_{\ell_\parallel}(z)$. Implementation of this procedure on the same simulations data as above yields the results plotted in Fig.\ref{fig:my_anisotropic2D}. Run I ($B_0=0$, Fig.\ref{fig:my_anisotropic2D}(a) ) shows an almost isotropic behaviour with only a slightly weaker cascade rate in the $z$ direction, while in run II ($B_0=2$,  Fig.\ref{fig:my_anisotropic2D}(b)) the cascade develops preferentially in the perpendicular plane with a very weak dependence on $\ell_\parallel$.
 
The results of Fig.\ref{fig:my_anisotropic2D} showing the anisotropic energy transfers when $B_0 \neq 0$, are in agreement with theoretical expectation for MHD and HMHD turbulence. However, to the best of our knowledge, this is the first time that a full 3D anisotropic cascade rate is directly revealed, while previous studies dealt with energy spectral anisotropy \cite{meyrand_anomalous_2013,cho_2000}, which is rather a consequence of the anisotropic cascade shown in Fig.\ref{fig:my_anisotropic2D}. It is worth noticing that a similar plot of anisotropic cascade is obtained experimentally in rotating fluid turbulence \citep{lamriben_direct_2011} 

\section{The cascade equation for the small scale energy}

It is possible to derive an equation complementary to Eq. \eqref{eq:MHD_en_fil_tot} that describes the time evolution of the energy density contained in scales smaller than the filtering scale $\l$. This sub-scale energy density is defined as $h_\ell=\left[\tau_\ell(u_i,u_i)+\tau_\ell(b_i,b_i)\right]/2$ (see \cite{aluie_coarse-grained_2017}) and its time evolution is given by (see Appendix \ref{ap:sm_sc}):
  \begin{equation}
  \begin{split}
     \partial_t h_\ell=& -\nabla\cdot\mathbfcal{J}^{\text{LS}}-\nabla\cdot\mathbfcal{J}^{\text{SS}}-\nabla\cdot\mathbfcal{J}^{\text{p}}\\&+\pi_\ell+\tau_\ell(u_i,d_{\nu,i})+\tau_\ell(u_i,d_{\eta,i})+\tau_\ell(u_i,f_i)
     \label{eq:ss}
     \end{split}
 \end{equation}
 where all quantities are function of both the position $\x$ and the scale $\ell$.\\
In the first line we find  $\nabla\cdot\mathbfcal{J}^\text{LS},\nabla\cdot\mathbfcal{J}^\text{SS},\nabla\cdot\mathbfcal{J}^{\text{p}}$ (whose expressions are given in Appendix \ref{ap:sm_sc}), which describe the spatial transport of $h_\ell$ due to large and small scale fields and pressure interactions, respectively. The local in space cascade rate $\pi_\ell(\x)$  appears as a sink in the large scale equation \eqref{eq:MHD_en_fil_tot} and as a source here, its full expression is given by:
 \begin{equation}
 \begin{split}
     \pi_\ell(\x)=&- \partial_j\bar{u}_i(\tau_\ell(u_i,u_j)-\tau_\ell(b_i,b_j))-\epsilon_{\alpha\beta\gamma}\bar{j}_\alpha\tau(u_\beta,b_\gamma)\\
     &+d_i\varepsilon_{\alpha\beta\gamma}\bar{j}_\alpha\tau(j_\beta,b_\gamma)
 \end{split}
 \label{eq:pi_full}
 \end{equation}
  where we introduced the Levi-Civita tensor $\varepsilon_{\alpha\beta\gamma}$, with the usual summation rule over repeated indices.
We stress here that $\pi_\ell(\x)$ is the only term able to exchange energy between the large and small scales  at a given position $\x$. The other terms instead  are associated either to spatial transport or to forcing/dissipation. These last two processes enter the small scale energy equation as a difference of filtered terms, \emph{e.g.} $\tau_\ell(u_i,d_{\nu,i})=\overline{(\bm{u}\cdot\bm{d}_\nu)}_\ell-\bar{\bm{u}}_\ell\cdot\bar{\bm{d}_\nu}_\ell$. To aid in the physical interpretation we show (see Appendix \ref{ap:tau}) that by averaging over a spatial region of characteristic size $L\gg \ell$ we recover the relation
 \begin{equation*}
     \langle \tau(f,g) \rangle_L\approx\langle f'g'\rangle_L
 \end{equation*}
 where $f'=f-\bar{f}, g'=g-\bar{g}$ are the ``unresolved" (subscale) fluctuations.
In this view, we can interpret the forcing/dissipation terms in equation \eqref{eq:ss} as the contributions to these processes coming from scales $<\ell$. For this reason we can write:
\begin{equation}
\begin{array}{c}
\langle\tau_\ell(u_i,f_i)\rangle_L\approx \langle\bm{u}'\cdot\bm{f}'\rangle_L\approx 0\\
\langle\tau_\ell(u_i,d_{\nu,i})+\tau_\ell(u_i,d_{\eta,i})\rangle_L\approx\langle \u'\cdot\bm{d}_\nu'+\b'\cdot\bm{d}_\eta'\rangle_L   
\end{array}
\label{eq:diss}
\end{equation}
where in the first line we stated that the forcing injects energy at large scales only and in the second line we recover the contrtibution of dissipation due to scales smaller than $\ell$.

  \section{Cascade and Local Dissipation}\label{sec:diss_proxy}
One of the main results of the Kolmogorov theory of turbulence is that the mean cascade rate (\emph{i.e.} averaged over the whole simulation box) is representative of the dissipation rate. We want to show that this holds even when averaging over much smaller spatial regions. In particular, we will prove that at small filtering scales $\ell$, and under some assumptions, the local cascade rate and the local dissipation rate match \emph{quasi-locally}. In other words, the amount of energy cascading across the (small) scale $\ell$ at position $\x$ will be eventually dissipated at close locations. Choosing a region of characteristic size $L$, we will derive the smallest $L$ for which the local cascade rate and the local dissipation match when spatially averaged over such region.\\
More precisely, we will prove that if the region size $L$ satisfies the two inequalities $L\gg\ell$,$L\gg\ell B_0/\delta b$ (the latter becoming redundant in strong turbulence with $\delta b/B_0>1$) then the following relation holds:

 \begin{equation}
     \langle \pi_\ell(\x)\rangle_L\approx -\langle\u'\cdot\bm{d}'_\nu+\b'\cdot\bm{d}'_\eta \rangle_L=\langle\varepsilon_{diss}^{<\ell}(\x)\rangle_L
     \label{eq:main}
 \end{equation}
where on the right side of Eq. \eqref{eq:main} we obtain the \emph{local} dissipation due to scales smaller than $\ell$ averaged on a region of size $L$.
We stress that the size of the region is not fixed and can be varied at will. In general we expect the agreement to get better as  $L\to L_{box}$, eventually recovering the ``global" result of Section \ref{sec:global}, however, we will show that this relation holds for regions of smaller size, effectively allowing us to study (quasi-)locally the processes of cascade and dissipation. 
 
 The starting point in deriving equation \eqref{eq:main} is the small scales ($\lesssim\ell$) equation \eqref{eq:ss}. For each position $\x$, we average over a region $\chi(\x,L)$, centered on $\x$, of characteristic size $L$ and of volume $V(L)\sim L^3$. For each quantity $q$ we can write $$\langle q \rangle_L=V(L)^{-1}\int _{\chi(\x,L)} q(\r)d\r.$$ This operation (\emph{de-facto} a new CG operation) is linear and as such it commutes with space and time derivatives. We can therefore write:
  \begin{equation}
     \partial_t \langle h_\ell\rangle_L=-\nabla\cdot\langle\mathbfcal{J}\rangle_L+\langle\pi_\ell\rangle_L +\langle\tau_\ell(u_i,d_{\nu,i})+\tau_\ell(b_i,d_{\eta,i})\rangle_L
     \label{eq:sm_sc_21}
 \end{equation}
where $\mathbfcal{J}=\mathbfcal{J}^{LS}+\mathbfcal{J}^{SS}+\mathbfcal{J}^{p}$. We can use this equation to study the characteristic time scales associated with the different terms.\\
We recall that there are two distinct scales: $\ell$ is the filtering scale across which energy cascades; $L$ is the size of the spatial region upon which we compute the spatial average. The filtering scale $\ell$ helps us to introduce two characteristic velocities: the large (at scales $>\ell$) and small ($<\ell$) scale velocity $\bar{u},u'$, respectively. The same filtering can be applied for the magnetic field with the only difference that the mean field cannot be removed by a Galilean transformation. However, we can decompose the characteristic large scale field as $\bar{b} = B_0+\delta\bar{b}$. The small scale field $\b' $is not affected by $B_0$. We furthermore assume the following Alfv\'enic ordering: 
$$\bar{u}\sim \delta\bar{b}, \quad u'\sim b'
$$
and denote these quantities as $\bar{U},U'$, which apply indistinctly to $u$ and $b$. 
Furthermore, we use $\bar{\bm{j}}\sim \delta\bar{b}/\ell, \bm{j}'\sim b'/\ell$ and, more importantly, $\tau(f,g)\sim f'g', \tau(f,g,h)\sim f'g'h'$ (see Appendix \ref{ap:tau} for the justification of the latter).
 \subsection{Nonlinear cascade}
Let us consider $\pi_\ell$, the local in space cascade rate. By spatially averaging $\pi_\ell$, see Eq. \eqref{eq:pi_full}, over a region of size $L$ we get:
 \begin{equation}
 \begin{split}
     \langle\pi_\ell\rangle_L=&-\langle \partial_j\bar{u}_i(\tau_\ell(u_i,u_j)-\tau_\ell(b_i,b_j))+\varepsilon_{\alpha\beta\gamma}\bar{j}_\alpha\tau(u_\beta,b_\gamma)\rangle_L\\
     &+d_i\langle \varepsilon_{\alpha\beta\gamma}\bar{j}_\alpha\tau(j_\beta,b_\gamma)\rangle_L
     \end{split}
 \end{equation}
It is straightforward to show that the order of $\langle\pi_\ell(\x)\rangle_L$, denoted $[\langle\pi_\ell\rangle_L]$ is given by:
 \begin{equation*}
     [\langle\pi_\ell\rangle_L]=\frac{\bar{U}}{\ell} U'^2+\frac{d_i}{\ell}\frac{\bar{U}}{\ell} U'^2
 \end{equation*}
so that the two characteristic times of the cascade can be found by computing $[\langle h_\ell\rangle_L]/[\langle\pi_\ell\rangle_L]$: $$t_{\text{NL}}^\text{MHD}=\frac{\ell}{\bar{U}}, \qquad t_{\text{NL}}^{Hall}=t_{\text{NL}}^\text{MHD}\frac{\ell}{d_i}$$
These quantities describe how fast energy cascades across scale $\ell$. In particular we recover the \emph{eddy turn-over time} at scale $\ell$ as the characteristic time of nonlinear cascade.
 
\subsection{Large scale spatial transport}
The terms governing the spatial transport of $\langle h_\ell \rangle_L$ due to the large scale fields read (see Appendix \ref{ap:sm_sc}):
\begin{equation}
\begin{split}
    &\nabla\cdot\langle\mathbfcal{J}^\text{LS}\rangle_L=\nabla\cdot \langle h_\ell\bar{u}_j-\tau(u_i,b_i)\bar{b}_j\rangle_L\\ &d_i\nabla\cdot\langle\bar{b}_j\tau(b_i,j_i)+\bar{j}_i\tau(b_i,b_j)-\bar{j}_j\tau(b_i,b_i)-\bar{b}_i\tau(j_j,b_i)\rangle_L
    \end{split}
\end{equation}
The divergence acts on quantities averaged over a region of size $L$. Therefore the characteristic scale of the spatial derivatives is $1/L$ as all fluctuations with smaller scales have been removed. With this in mind we can proceed to derive the characteristic time by which each of these processes extract or bring energy inside this region of size $L$: 
$$
    t_{\text{LS}}=\frac{L}{\bar{U}}, \quad t_{A}=\frac{L}{B_0}, \quad t_{\text{LS}}^{Hall}= t_{\text{LS}}\frac{\ell}{d_i}, \quad t_{\text{Hall}}=t_A\frac{\ell}{d_i}
$$
Alongside the characteristic time of energy advection by the large scale flow $t_\text{LS}$ we recognize the linear Alfv\'en time $t_A$ that represents the energy transport due to propagating Alfv\'en waves at scale $L$. Additionally, faster modes that have a time scale $t_{Hall}=t_A\ell/d_i$ exist as the scales approach $d_i$, which can be identified as whistler waves with a dispersion relation $\omega \sim k_{\parallel}kB_0 d_i$ \citep{sahraoui_waves_2007}.

\subsection{Small scale transport}

The effect of the small scale fluctuations in the spatial transport of $\langle h_\ell \rangle_L$ is given by (see Appendix \ref{ap:sm_sc}):
\begin{equation}
\begin{split}
    \nabla\cdot\langle\mathbfcal{J}^{\text{SS}}\rangle_L&=\nabla\cdot\left\langle \frac{\tau(b_i,b_i,u_j)+\tau(u_i,u_i,u_j)}{2}-\tau(u_i,b_i,b_j)\right\rangle_L\\
    &+d_i\nabla\cdot\left\langle\tau(b_i,b_j,j_i)-\tau(j_j,b_i,b_i)\right\rangle_L
    \end{split}
\end{equation}
The corresponding characteristic time scales read:
\begin{equation*}
    t_{\text{SS}}=\frac{L}{U'}, \quad t_{\text{SS}}^{\text{Hall}}=t_{\text{SS}}\frac{\ell}{d_i}
\end{equation*}

\subsection{Pressure term}
The last term that we need to analyze is the spatial transport one, $\nabla\cdot\langle\tau(u,P)\rangle_L$,  involving the total plasma pressure. We recall that in the incompressible HMHD model pressure is completely determined by $\u,\b$: taking the divergence of \eqref{eq:HMHD}, pressure is found by solving the Poisson equation:
\begin{equation*}
    \nabla^2P=\partial_i\partial_j (u_iu_j-b_ib_j)
\end{equation*}
the equation for $P'=P-\bar{P}$ is easily derived, whose ordering is given by 
\begin{equation}
    [P']= U'^2+\bar{U}U'
\end{equation}
which yields the following ordering of the transport term
\begin{equation}
    [\nabla\cdot\langle\tau(u,P)\rangle_L]=\frac{1}{L}P'U'\approx \frac{U'^2U'}{L} +\frac{\bar{U}U'^2}{L}
\end{equation}
and the corresponding characteristic time scales
\begin{equation*}
    t_\text{SS}=\frac{L}{U'},  \quad
    t_\text{LS}=\frac{L}{\bar{U}}
\end{equation*}
both of which were already found in the analysis of the large and small scale transport terms.

\subsection{The fastest dynamical process}\label{fast}
We want to show that we can choose the region size $L$ so that the cross-scale energy transfer in the region is faster than spatial transport across the region borders. To do so we analyze the time-scales associated with different processes.
For the sake of simplicity the comparison is limited to the MHD range where the Hall term can be neglected. Including the Hall term will not modify any of the conclusion of this study based on the observation that the Hall term modifies the time scales of the nonlinear cascade and transport by the same (small scale) factor $\ell/d_i$.
We consider the nonlinear time scale $t_{\text{NL}}^\text{MHD}$ as a reference to obtain:
\small
\begin{equation}\label{order}
    \frac{t_{\text{NL}}^\text{MHD}}{t_{\text{LS}}}=\frac{\ell}{L}; \quad \frac{t_{\text{NL}}^\text{MHD}}{t_{\text{SS}}}=\frac{\ell}{L}\frac{U'}{\bar{U}}; \quad  \frac{t_{\text{NL}}^\text{MHD}}{t_{\text{A}}}=\frac{\ell}{L}\frac{B_0}{\bar{U}}\sim \frac{\ell}{L}\frac{B_0}{\delta\bar{b}}
\end{equation}
\normalsize
we notice that when $\ell/L\ll1$ and $U'<\bar{U}$ (the latter is justified by a power-law decline of the fluctuations in MHD turbulence) the first two conditions in relations (\ref{order}) yield the ordering $t_{\text{NL}}^\text{MHD}\ll t_{\text{LS}},t_{\text{SS}}$.\\  

If we further assume high amplitude fluctuations with respect to the background field, \emph{i.e.}, $\delta b/B_0\gtrsim1$, the condition $\ell/L\ll 1$ automatically implies the ordering $t_{\text{NL}}^\text{MHD}\ll t_{\text{A}}$. This should not be confused with the critical balance ordering \citep{goldreich_toward_1995} as the quantity $t_A$ is the time that an Alfv\'en wave takes to propagate over a distance $L$: as $L$ can be varied at will, $t_A$ can be made as large or as small as needed. In the general case the stronger the mean magnetic field compared to the fluctuations, the larger we should choose $L$ to maintain $t_{\text{NL}}^\text{MHD}\ll t_{\text{A}}\rightarrow L\gg \ell B_0/\delta b$ for a given $\ell$. 

To summarize, under the conditions $\ell/L\ll 1$ and $\ell/L\ll\delta b/B_0$, the latter becoming redundant in strong turbulence, $t_\text{NL}^\text{MHD}$ becomes the fastest time of the averaged dynamics. The interpretation is the following: considering a spatial region of characteristic size $L$, a certain amount of energy cascades across scale $\ell$  inside this region on time scales much faster than what it takes for the same amount of energy to be spatially transported across the region surface by large and small scale processes including linear (Alfv\'en) waves. \\

For the aforementioned reason, in the averaged small scale equation the spatial transport terms are slower than the nonlinear cascade. At a small filtering scale $\ell$ we anticipate the averaged cascade to be balanced by the averaged dissipation: as energy cascades in a given region it does not have time to be spatially transported outside the region before it is dissipated (note that this would imply some kind of \emph{short-time} stationarity of the small scale, averaged, energy $\langle h_\ell\rangle_L $).\\
Therefore, in equation \eqref{eq:sm_sc_21} we can expect the two fast processes to match:
\small
\begin{equation}\langle\pi_\ell(\x)\rangle_L\approx-\langle\tau_\ell(u_i,d_{\nu,i})+\tau_\ell(b_i,d_{\eta,i})\rangle_L\approx-\langle\u'\cdot\bm{d}'_\nu+\b'\cdot\bm{d}'_\eta \rangle_L\end{equation}
\normalsize
This equation shows, as stated at the beginning of this section, that the (quasi-)local cascade matches the contribution to the (quasi-)local dissipation coming from scales $\lesssim \ell$.\\
The result is general and does not require the existence of an inertial range as it does not involve the total dissipation, but only a part of it. However, if we assume the dissipation rate due to the large scales (\emph{i.e.} $\gtrsim \ell$) to be negligible, it can be further refined to obtain 
\begin{equation}
    \langle\pi_\ell(\x)\rangle_L\approx-\langle \u'\cdot\bm{d}'_\nu+\b'\cdot\bm{d}'_\eta \rangle_L\approx -\langle \u\cdot\bm{d}_\nu+\b\cdot\bm{d}_\eta \rangle_L
    \label{eq:main_def}
\end{equation}
which shows the approximate (quasi-)local balance between the cascade rate and total dissipation. This result will now be tested numerically.
\subsection{Numerical Validation}
\begin{figure}[]
    \centering
    \includegraphics[width=0.5\textwidth]{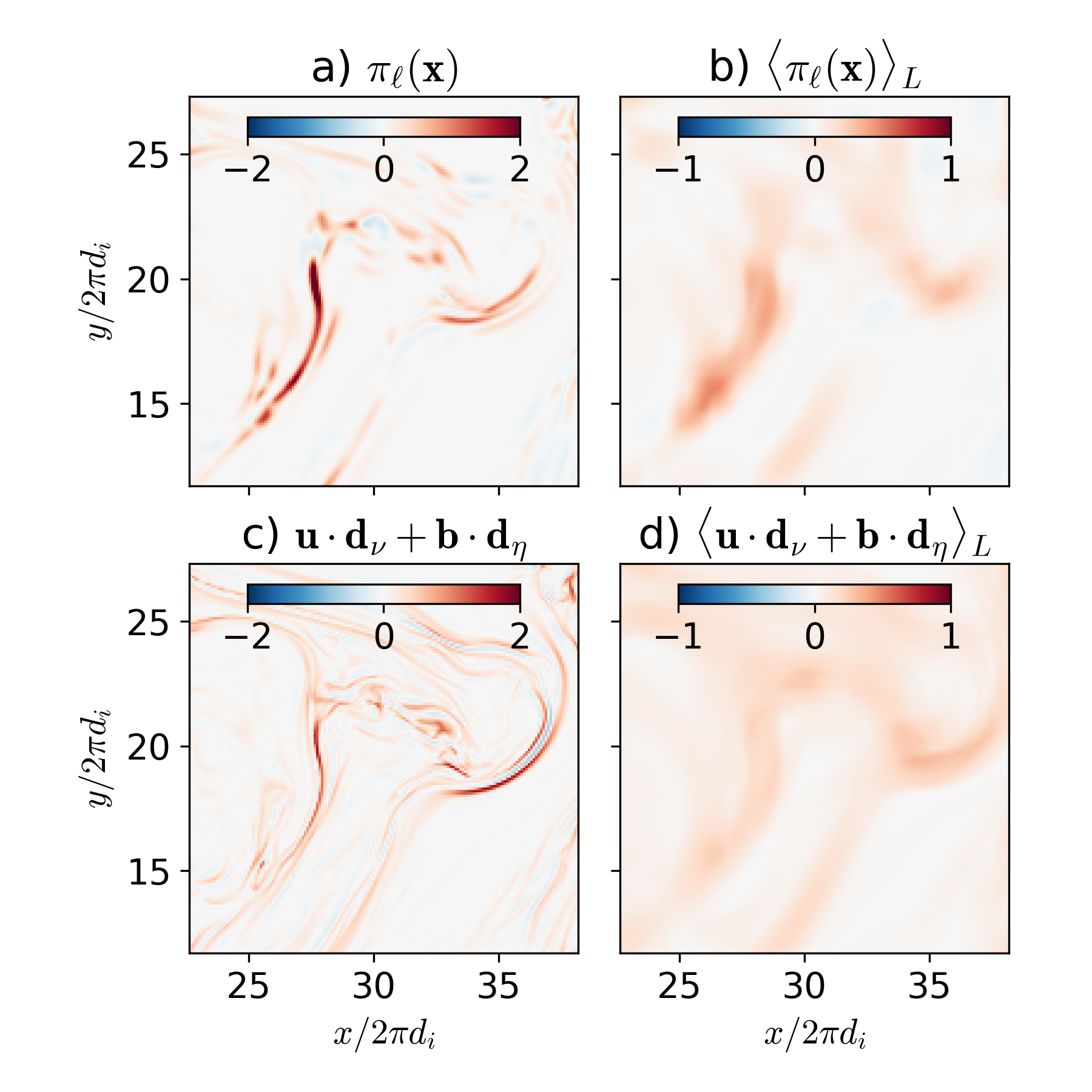}
    \caption{2D plots of the cascade rate $\pi_\ell$ across scale $\ell=0.3d_i$ (a) and dissipation (b) based on the data from run I. Figures (c,d) are obtained by computing for each point $\bm{x}$ the average on a cubic region of size $L=1.4d_i$ centered on $\x$.}
    \label{fig:local_plot}
\end{figure}
\begin{figure*}[]
    \centering
    \includegraphics[width=0.9\textwidth]{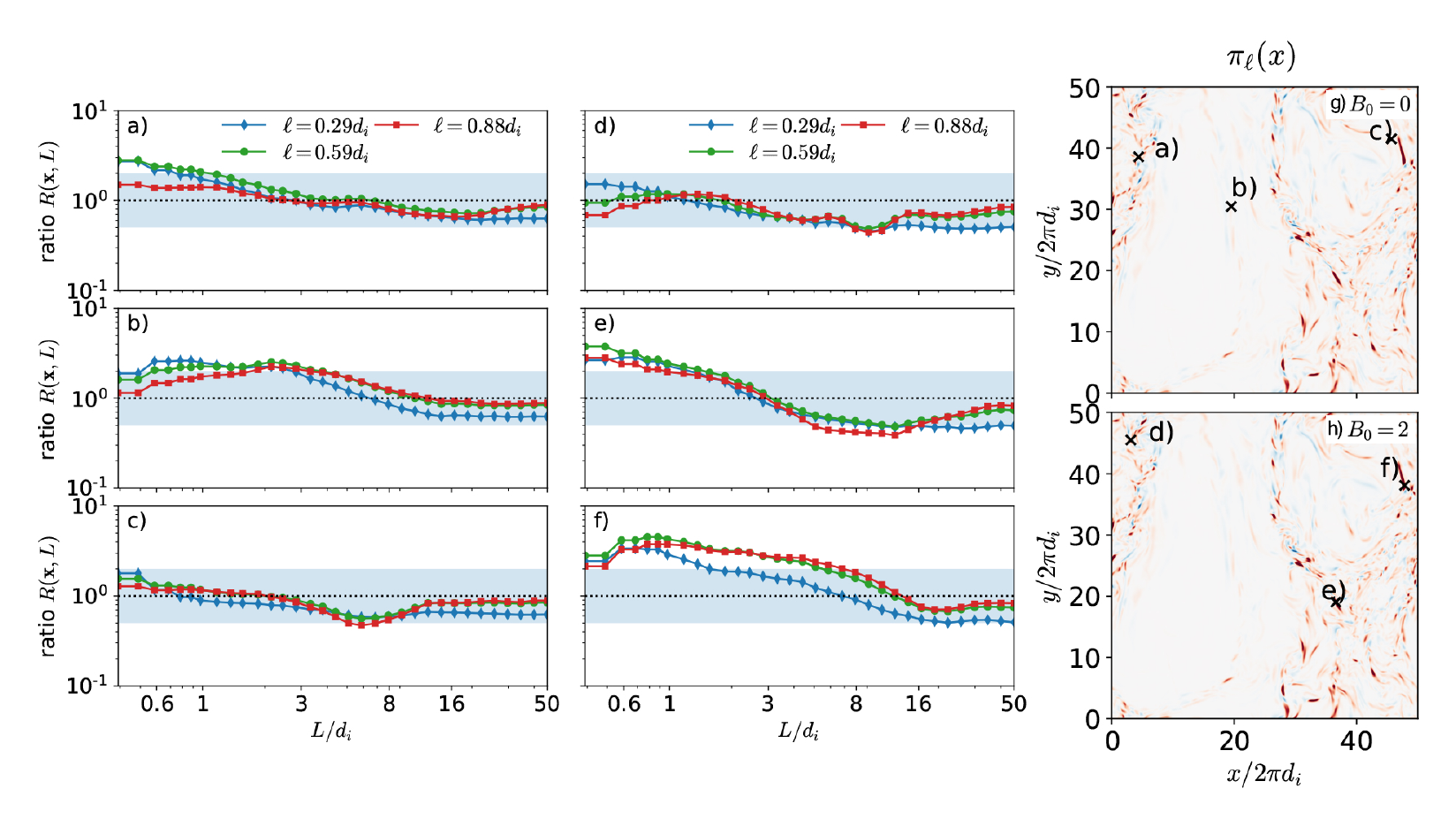}
    \caption{Ratio $R(\x,L)=-\langle\pi_\ell\rangle/\langle\u\cdot\bm{d}_\nu+\b\cdot\bm{d}_\eta \rangle$ between the local cascade rate and the local dissipation rate averaged on a cubic region of size $L$ centered around a given point $\x$. The results for three different points on the plane $z=25\times 2\pi d_i$ in run I (a-c) and run II (d-f) are shown. The shaded region individuates the interval $0.5<R<2$. }
    \label{fig:my_scaling}
\end{figure*}
A first glimpse on the local (in space) behavior of the cascade and dissipation at the filtering scale $\ell=0.3d_i$ is given in Fig. (\ref{fig:local_plot}(a) and (c)) based on the data from run I. Overall the two quantities exhibit similar patterns, indicating an approximate balance between them, although the extrema of the cascade rate can be larger locally by a factor $\sim 2$. We also observe that cascade rate $\pi_\ell(x)$ presents both positive and negative values. This is a clear sign that the nonlinear interactions work in both directions, bringing energy from large to small scales and \emph{vice-versa}, albeit at this small filtering scale $\ell=0.3d_i$ most of the energy is going towards the small scales and $\pi$ is mostly positive. Nevertheless, when integrated over a cubic region of size $L=1.4d_i$ centered on each point $\x$ (panel \ref{fig:local_plot}(b)), we recover a nearly positive flux, i.e. $\langle\pi_\ell(\x)\rangle_L>0$, a sign that on average the quasi-local turbulent cascade is {\it direct} since energy is carried towards the small scales. The same observations can be made about dissipation (panels \ref{fig:local_plot}(c) and (d)), which is indeed positive definite only when no energy leakage through the boundaries is assumed.

A thorough analysis can be done by pinpointing individual locations in space $\x_0$ that correspond to intense local transfers on which we can test the balance between the energy cascade and dissipation given by relation (\ref{eq:main_def}). The chosen locations are indicated in Fig.\ref{fig:my_scaling}  by labels (a)-(c) for run I and (d)-(f) for run II. Varying $L$, the size of the region over which the integration is performed, we are able to gauge $R(\x_0,L)$, the ratio between $\langle\pi_\ell\rangle_L$ and $-\langle\u\cdot\bm{d}_\nu+\b\cdot\bm{d}_\eta\rangle_L$.
The results in Fig.\ref{fig:my_scaling} show that the ratio between the two terms is close to 1 not only when $L=50d_i=L_\text{box}$ but it remains constant for a very large range of scales all the way down to $\ell \sim 8 d_i$.
This is consistent with relation \eqref{eq:main_def} which holds when $\ell/L\ll1$. It is however remarkable to notice that even for smaller values $\ell\sim L$ the ratio between the two terms remains comparable with 1. In particular, Fig.\ref{fig:my_scaling} shows that $0.5\leq R \leq 2$ at almost all scales for all the three points under study in run I. The same study for run II shows an overall good agreement between the local cascade and dissipation magnitude. However, the matching is less good than in run I, as the ratio $R(\x_0,L)$ departs from the reference value $1$ (i.e., perfect balance) at larger size box $L$ for run II than run I. This is likely to be caused by the presence of the mean magnetic field $B_0=2$ in run II, which introduces (large scale) Alfv\'en waves that spatially transport energy on comparable time scales than those of the nonlinear cascade and dissipation as discussed in Section \ref{fast}. Said differently, for run I the time scale ratio 
$t_{\text{NL}}^\text{MHD}/t_{\text{A}}\sim \ell B_0/(L\langle\delta \bar{b}\rangle_L)\sim 0$ since $B_0=0$ (i.e., $t_A \to \infty$) regardless of how large is the ratio $\ell/L$. This could explain the very good matching between the cascade and dissipation rates in run I even for very small integrated regions, i.e., $\ell/L\sim 1$ as observed in Fig.\ref{fig:my_scaling} (a)-(c).  
In run II, with $B_0=2$, the linear transport time scale $t_A $ becomes finite and can be comparable to $t_{\text{NL}}^\text{MHD}$ even for small ratios of $\ell/L$. In the limit case of a very strong mean field, $\delta b/B_0\ll 1$ the linear Alfv\'en time will become faster than the nonlinear time at all scales, and relation \eqref{eq:main_def} would no longer hold.\\
\begin{figure*}[]
    \centering
    \includegraphics[width=0.8\textwidth]{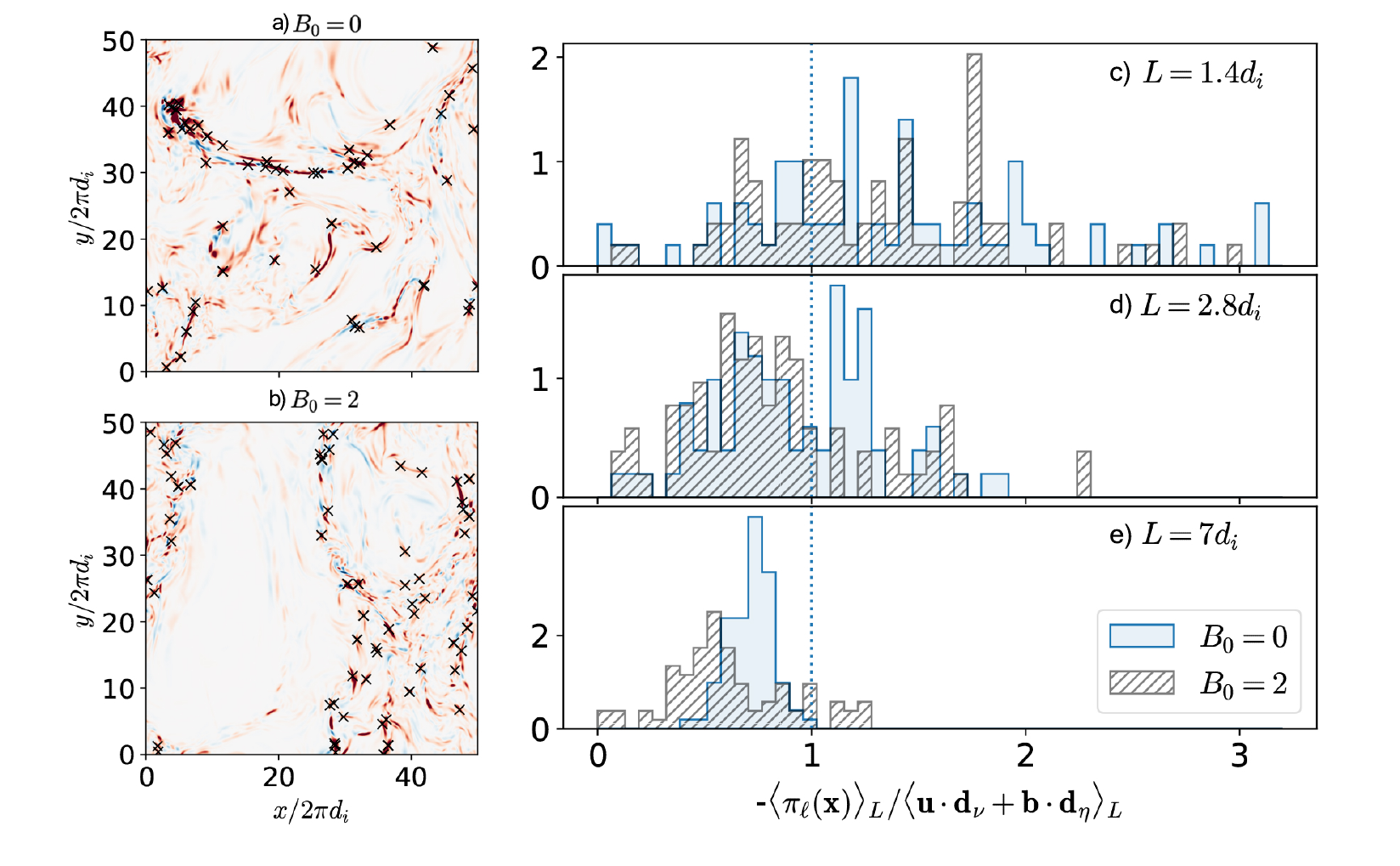}left
    \caption{The normalized distributions of the ratio $-\langle\pi_\ell(\mathbf{x})\rangle_L/\langle \mathbf{u}\cdot\mathbf{d}_\nu+\mathbf{b}\cdot\mathbf{d}_\eta\rangle_L$ (c-e) computed at scale $\ell=0.3 d_i$ and plane $z=25\times 2\pi d_i$ in the regions of intense cross-scale transfer marked by crosses for runs I (a) and II (b). Three different sizes $L$ of the integrated regions are considered. Filled, colored values correspond to run I ($B_0=0$), while black, hatched values to run II $B_0=2$.}
    \label{fig:my_scaling_statistic}
\end{figure*}

To obtain a more complete picture of the balance between the quasi-local cascade and dissipation we performed the same study over a larger sample of locations shown in Fig \ref{fig:my_scaling_statistic}. For both runs we consider some of the local maxima of $\pi_\ell(\x)$, $\ell=0.3 d_i$ in the plane $z=25\times2\pi d_i$, for each of these points the ratio $R(\x_0,L)$ is computed and its histogram is shown for each size $L$ of the integrated box. As expected, for both runs we observe that as $L$ decreases the width of the histogram (i.e., dispersion) increases, and its mean tends to shift to larger values. This indicates that, on average, the local cascade tends to slightly dominate the local dissipation. We also observe that dispersion around the reference value $R(\x_0,L) \sim 1$ (perfect balance) is more prominent in run II than run I, which confirms the role of the background magnetic field $B_0$ in shortening the linear transport time, thus competing with the pair cascade-dissipation discussed above.

\section{Conclusions}
In this paper we have derived the \emph{large} and \emph{small} scales CG equations for incompressible HMHD and showed that the quantity $\pi_\ell(\x)$ is the local (at position $\x$) energy transfer across the scale $\ell$. This quantity, when averaged over the simulation domain allows us to recover results consistent with the so-called ``third-order" laws. It represents an improvement over the current state-of-the-art in this field as it does not rely on the assumptions of ergodicity or the existence of the inertial range. The model is further generalized to account for anisotropic cascade in the presence of a background magnetic field. The major strength of this new theory is to provide a means of estimating local energy dissipation in turbulent plasmas. This is shown to be achieved at a cost of fairly loose assumptions, the main ones being the scale separation between the filtering scale $\ell$ and the integration region size $L$, moderate-to-large amplitude of the turbulent fluctuations with respect to the background field $B_0$. This last condition is expected in strong turbulence. These two assumptions are required to minimize the role of the spatial transport across the region of integration, enforcing thus the balance between local nonlinear cascade and dissipation.   

The theory was tested successfully on two simulations featuring different intensities of the background field $B_0$. In agreement with the time scale estimate of the various processes, a very good local balance between cascade and dissipation is found in run with $B_0=0$ even when $L\sim \ell$, while a moderate imbalance between the two is observed at larger values $L$ for run with $B_0=2$. We conjecture that this behaviour is due to the role of Alfv\'en waves that spatially transport energy across the integration region on time scales comparable to those of the cascade and dissipation.

An immediate application of this novel theory is to estimate energy dissipation in localized structures (e.g., reconnecting current sheets) frequently reported in numerical simulations \citep{arro_statistical_2020} and spacecraft observations in the near-Earth space \citep{chasapis_thin_2015,huang2018}. Indeed, even if the theory is based on a fluid (HMHD) model, the fact that $\pi_\ell(\x)$ is shown to reflect local dissipation (within the aforementioned assumptions) regardless of the explicit form of the dissipation operators makes it particularly relevant to collisionless plasmas where the damping of the fluid and electromagnetic fluctuations is believed to originate from kinetic processes that would show up in the moment equation as complex damping operators $\bm{d}_\nu,\bm{d}_\eta$.

The present theory can be extended in various directions. For instance to the two-fluid model where electron inertia can be accounted for (particularly relevant in reconnection studies and small scale turbulence, see for instance \citep{faganello_being_2009} and references therein), or by incorporating compressible effects \citep{eyink_cascades_2018-1, eyink_cascades_2018}. Regarding its implementation on simulations and spacecraft data, an integration over an anisotropic domain $L_\perp\ll L_\parallel$ would be more appropriate to capture the difference in time scale of the various processes along the two directions. Lastly, in numerical simulation an integration over the actual 3D volume of the structures (e.g., current sheets, plasmoids) would be a significant improvement with respect to the cubic box adopted here. 

 \appendix

\section*{Acknowledgement}
The authors thank Prof. P. Mininni and the developers of the GHOST code for providing the code used to run the simulations presented in this work.\\
This work was granted access to the HPC resources of CINES under allocation 2021 A0090407714 made by GENCI. It is supported by the CNRS/CONICET Laboratoire International Associé (LIA) MAGNETO.

 \section{On the generalized central moments } \label{ap:tau}
When studying turbulence using the statistical approach the original field $u$ is decomposed into its (ensemble) average part $\langle\u\rangle$ and a fluctuations $\u'$ whose sum gives the original quantity $\u=\langle\u\rangle+\u'$. In this framework we can derive relations concerning the central moments of the fluctuations $\langle u'_iu'_j\rangle,\langle u'_iu'_ju'_k\rangle$,\dots. It can be shown \citep{frisch_turbulence_1995, monin_statistical_2013} that:

\begin{equation}
\begin{split}
    &\langle u'_iu'_j\rangle=\langle u_iu_j\rangle-\langle u_i\rangle\langle u_j\rangle\\
    &\langle u'_iu'_ju'_k\rangle=\langle u_iu_ju_k\rangle-\langle u_i\rangle\langle u_j'u_k'\rangle -\langle u_j\rangle\langle u_i'u_k'\rangle\\
    & \qquad\qquad\qquad  -\langle u_k\rangle\langle u_i'u_j'\rangle -\langle u_i\rangle\langle u_j\rangle\langle u_k\rangle\\
    &\langle u'_iu'_ju'_ku'_l\rangle=\dots
\end{split}
\label{eq:ap_moms}
\end{equation}
When using the filtering approach (CG), we decompose the field into its large scale $\bar{\u}$ and ``unresolved" $\u-\bar{\u}$ contributions, replacing the ensemble average $\langle\cdot\rangle$ with the CG operation. To recover a set of relations equivalent to \eqref{eq:ap_moms} the seminal work of M.Germano \cite{germano_1992} introduces the \emph{generalized central moments} $\tau(u_i,u_j),\tau(u_i,u_j,u_k),\dots$, formally defined as:
\begin{equation}
\begin{split}
    &\tau(u_i,u_j)=\overline{u_iu_j}-\bar{u}_i\bar{u}_j\\
    &\tau(u_i,u_j,u_k)=\overline{u_iu_ju_k}-\bar{u}_i\tau(u_j,u_k) -\bar{u}_j\tau(u_i,u_k)\\
    & \qquad\qquad\qquad  -\bar{u}_k\tau(u_i,u_j) -\bar{u}_i\bar{u}_j\bar{u}_k\\
    &\tau(u_i,u_j,u_k,u_l)=\dots
\end{split}
\end{equation}
which are employed in this work and allow us to recover simple filtered equations. To aid in the physical interpretation some useful relations can be derived to link the generalized central moments to the field fluctuations \citep{vreman_realizability_1994}. We start with:
\small
\begin{equation}
\begin{split}
    &\tau(f,g)=\overline{fg}-\bar{f}\bar{g}=\left(\overline{fg}-\bar{f}\bar{g}\right) +\left(-\bar{f}\bar{g}+\bar{f}\bar{g}\right)\\
    =& \int d\r G_\ell(\x+\r)f(\x+\r)g(\x+\r)-\bar{f}(\x)\int d\r G_\ell(\x+\r)g(\x+\r)+\\
    -&\bar{g}(\x)\int d\r G_\ell(\x+\r) f(\x+\r) +\bar{f}(\x)\bar{g}(\x)\int d\r G_\ell(\x+\r)\\
    =&\int d\r G_\ell(\x+\r)\left[ f(\x+\r)-\bar{f}(\x)\right]g(\x+\r)-\bar{g}(\x)\times\\
    \times&\int d\r G_\ell(\x+\r)\left[f(\x+\r)-\bar{f}(\x)\right] =\\
    =&\int d\r G_\ell(\x+\r)\left[f(\x+\r)-\bar{f}(\x)\right]\left[g(\x+\r)-\bar{g}(\x)\right]
\end{split}
\label{eq:vreman}
\end{equation}
\normalsize
Performing a second order Taylor expansion:  
\[
f(\x+\r)= f(\x)+r_j\frac{\partial f}{\partial x_j}+\frac{r_ir_j}{2}\frac{\partial^2 f}{\partial x_i\partial x_j}+o(|\r|^2)
\]
and denoting the turbulent, subscale fluctuations with $f'=f-\bar{f}$, we have:
\begin{equation*}
\begin{split}
    \tau(f,g)\sim& f'(\x)g'(\x)+\left[g'\nabla f +f'\nabla g\right]_{\x}\cdot\int d\r \r G_\ell(\r)\\
    +&\left[\frac{\partial f }{\partial x_i}\frac{\partial g }{\partial x_j}+\frac{\partial f }{\partial x_j}\frac{\partial g }{\partial x_i}\right]\int d\r G_\ell(\r)r_ir_j\\
    +&\left[g'\frac{\partial^2 f }{2\partial x_i\partial x_j}+f'\frac{\partial^2 g }{2\partial x_i\partial x_j}\right]\int d\r G_\ell(\r)r_ir_j
    \end{split}
\end{equation*}
Since $\int d\r G_\ell(\r)\r=0$, and the second order centered moments matrix $\int d\r G_\ell(\r)r_ir_j\sim \ell^2 \delta_{ij}$ (with the Gaussian filter used), the relation
$\tau(f,g)\sim f'g'$ is valid up to the second order in $\ell/\Lambda$. Here $\Lambda^{-1}\sim \min\{\nabla f/f,\nabla g/g\}$ is the characteristic scale of the field gradients. The parameter $\ell/\Lambda$ is however not guaranteed to be small as the fields have a Fourier spectrum that extends all the way down to the dissipation scales. Therefore, an even more precise relation can be obtained by performing another CG operation, or, in general, by averaging over a spatial region of size $L\gg\ell$:
 \begin{equation}
     \langle\tau(f,g)\rangle_L= \langle f'g'\rangle_L+\mathcal{O}\left((l/\Lambda')^2\right) 
     \label{eq:fluct}
 \end{equation}
where  $\Lambda'$, the gradient characteristic scale of the averaged fields, is of order $L$ since the spatial average removes fluctuations  at smaller scales.\\

A similar result can be derived for the third order generalized moment defined as:
\begin{equation}
    \tau(f,g,h)=\overline{fgh}-\bar{f}\bar{g}\bar{h}-\bar{f}\tau(g,h)-\bar{g}\tau(f,h)-\bar{h}\tau(f,g)
\end{equation}
a direct computation, denoting $\x'=\x+\r$ shows:

\small
\begin{equation}
\begin{split}
    &\int d\r G_\ell(\x') \left[f(\x')-\bar{f}(\x)\right]\left[g(\x')-\bar{g}(\x)\right]\left[h(\x')-\bar{h}(\x)\right]=\\
    =& \overline{fgh}-\bar{f}\bar{g}\bar{h}+\, [-\bar{f}\overline{gh}+\bar{f}\bar{g}\bar{h}]+ [-\bar{g}\overline{fh}+\bar{g}\bar{f}\bar{h}]+ [-\overline{fg}\bar{h}-\bar{f}\bar{g}\bar{h}] =\\
    =& \overline{fgh}-\bar{f}\bar{g}\bar{h} \qquad - \bar{f}\tau(g,h) \qquad -\bar{g}\tau(f,g)\qquad-\bar{h}\tau(f,g)=\\&=\tau(f,g,h)
    \end{split}
\end{equation}
\normalsize
and, as for the second order generalized central moment, a Taylor expansion followed by a spatial average leads to
\begin{equation}
    \langle\tau(f,g,h)\rangle_L=\langle f'g'h'\rangle_L+\mathcal{O}((\ell/L)^2)
\end{equation}
which is precisely the relation used in the main text.

 \section{DERIVATION OF THE SMALL SCALE EQUATIONS} \label{ap:sm_sc}
 The energy contained at sub-filtering scales $<\ell$ is defined as  $h_\ell=\left[\tau_\ell(u_i,u_i)+\tau_\ell(b_i,b_i)\right]/2$ and is a positive quantity at every point in space in virtue of relation \eqref{eq:vreman}, \emph{e.g.} $\tau(u_i,u_i)=\int d\r G_\ell(\x+\r)|u_i(\x+\r)-\bar{u}_i(\x)|^2>0$. Furthermore when integrating over the spatial domain we recover the total sub-scale energy $\int d^3\x h_\ell(\x)=\int d^3x \left[ (|\u|^2-|\bar{\u}_\ell|^2)+ (|\b|^2-|\bar{\b}_\ell|^2) \right]/2$, (see \cite{aluie_coarse-grained_2017}).\\
 To compute the evolution of $h_\ell$ we filter on a scale $\ell$ the energy equation and we subtract the equation for the large scale energy $(|\bar{\u}|^2+|\bar{\b}|^2)/2$. We readily obtain for kinetic and magnetic energy densities:
 \begin{widetext}
 \begin{equation}
\begin{split}
    \partial_t \frac{\tau(u_i,u_i)}{2}=&-\partial_j\left[\frac{\tau(u_i,u_i)}{2}\bar{u}_j+\frac{\tau(u_i,u_i,u_j)}{2}+\tau(u_j,P)-\tau(u_i,b_i,b_j)-\tau(u_i,b_j)\bar{b}_i-\tau(u_i,b_i)\bar{b}_j\right]\\
    &+\pi^u+(\bar{b_i}\bar{b}_j\partial\bar{u}_j-\overline{b_ib_j\partial_j u_i})+\tau(u_i,d_{\nu,i})+\tau(u_i,f)
\end{split}
\end{equation}

\begin{equation}
\begin{split}
    \partial_t \frac{\tau(b_i,b_i)}{2}=&-\partial_j\left[\frac{\tau(b_i,b_i)}{2}\bar{u}_j+\frac{\tau(b_i,b_i,u_j)}{2}+\bar{b}_i\tau(u_i,b_j)\right]\\
    &-d_i\partial_j\left[\tau(b_i,b_j,j_i)-\tau(j_j,b_i,b_i)+\bar{b}_j\tau(b_i,j_i)+\bar{j}_i\tau(b_i,b_j)-\bar{j}_j\tau(b_i,b_i)-\bar{b}_i\tau(j_j,b_i)\right]\\
    &+\pi^b+(\overline{b_ib_j\partial_j u_i}-\bar{b_i}\bar{b}_j\partial\bar{u}_j)+\tau(b_i,d_{\eta,i})
\end{split}
\end{equation}
And taking the sum of the two equations we obtain
\begin{equation}
\begin{split}
    \partial_t \frac{\tau(u_i,u_i)+\tau(b_i,b_i)}{2}=&-\partial_j\left[\frac{\tau(u_i,u_i)+\tau(b_i,b_i)}{2}\bar{u}_j+\tau(u_j,P)-\tau(u_i,b_i)\bar{b}_j\right]
    -\partial_j\left[\frac{\tau(b_i,b_i,u_j)+\tau(u_i,u_i,u_j)}{2}-\tau(u_i,b_i,b_j)\right]\\
    &-d_i\partial_j\left[\tau(b_i,b_j,j_i)-\tau(j_j,b_i,b_i)+\bar{b}_j\tau(b_i,j_i)+\bar{j}_i\tau(b_i,b_j)-\bar{j}_j\tau(b_i,b_i)-\bar{b}_i\tau(j_j,b_i)\right]\\
    &+\pi^u+\pi^b+\tau(b_i,d_{\eta,i})+\tau(u_i,d_{\nu,i})+\tau(u_i,f)
\end{split}
\label{eq:sm_sc_app}
\end{equation}
\end{widetext}
We can group the contribution of the spatial transport of the small scale energy $h_\ell=\left[\tau(u_i,u_i)+\tau(b_i,b_i)\right]/2$ due to the large scale fields:
\begin{equation*}
\begin{split}
    &\nabla\cdot\mathbfcal{J}^\text{LS}=\nabla\cdot \left[h_\ell\bar{u}_j-\tau(u_i,b_i)\bar{b}_j\right]+\\ +&d_i\nabla\cdot\left[\bar{b}_j\tau(b_i,j_i)+\bar{j}_i\tau(b_i,b_j)-\bar{j}_j\tau(b_i,b_i)-\bar{b}_i\tau(j_j,b_i)\right]
    \end{split}
\end{equation*}

Similarly, the spatial transport of $h_\ell$ due to the subscale fields is given by:
\begin{equation*}
\begin{split}
    &\nabla\cdot\mathbfcal{J}^{\text{SS}}=\nabla\cdot\left[ \frac{\tau(b_i,b_i,u_j)+\tau(u_i,u_i,u_j)}{2}-\tau(u_i,b_i,b_j)\right]+\\
    +&d_i\nabla\cdot\left[\tau(b_i,b_j,j_i)-\tau(j_j,b_i,b_i)\right]
    \end{split}
\end{equation*}
and lastly the spatial transport due to pressure interactions:
\begin{equation*}
    \nabla\cdot\mathbfcal{J}^\text{p}=\nabla\cdot\tau(u_j,P)
\end{equation*}
so that equation \eqref{eq:sm_sc_app} can be written in compact form as:
\begin{equation}
\begin{split}
    \partial_t h_\ell=&-\nabla\cdot\left[\mathbfcal{J}^{LS}+\mathbfcal{J}^{SS}+\mathbfcal{J}^{p}\right]\\+&\pi_\ell+\tau(b_i,d_{\eta,i})+\tau(u_i,d_{\nu,i})+\tau(u_i,f)
    \end{split}
\end{equation}

\end{document}